# An Easily Constructed, Tuning Free, Ultra-broadband

# Probe for NMR

D. Murphree, S.B. Cahn, D. Rahmlow, D. DeMille

*Yale University Department of Physics, New Haven, CT 06520*

Dennis Murphree
SPL Room 25
217 Prospect St.
New Haven, CT 06520-8120

Email: dennis.murphree@yale.edu
Phone: 203-436-4437
Fax: 203-432-9710

# An Easily Constructed, Tuning Free, Ultra-broadband Probe for NMR

D. Murphree, S.B. Cahn, D. Rahmlow, D. DeMille
*Yale University Department of Physics, New Haven, CT 06520*

We have developed an easy to construct, non-resonant wideband NMR probe. The probe is of the saddle coil geometry and is designed such that the coil itself forms a transmission line. The probe thus requires no tuning or matching elements. We use the probe with a spectrometer whose duplexer circuitry employs a simple RF switch instead of the more common $\lambda/4$ lines, so the entire probe and spectrometer perform in an essentially frequency-independent manner. Despite being designed with electro- and magnetostatic formulas, the probe performs well at frequencies up to 150 MHz and beyond. We expect that with additional design effort, the probe could be modified for use at significantly higher frequencies. Because our construction method relies on commercial circuit fabrication techniques, identical probes can be easily and accurately produced.

# 1 Introduction

Most nuclear magnetic resonance measurements use resonant tuned probes for generation and detection of radiofrequency (RF) magnetic fields. However, a variety of applications require probes capable of wideband operation. Here we present the design for such a probe, measure its performance, and describe our experiences of constructing probes of this type.

The design of our probes has been driven by a specific experiment underway in our group, with the goal to study electroweak interactions inside molecules. This experiment requires the ability to shim and monitor a magnetic field in the range 0.1-0.6 T, to a homogeneity of 0.1ppm, with rapid adjustments to large changes in the field strength. In order to do so, we will use an array of thirty-two spatially distributed NMR probes to provide a precise real-time map of the magnetic field. Because we will not have physical access to the probes in order to tune them as we change the field, we need a probe which is non-resonant and automatically impedance matched for a broad range of frequencies. Although we are using the probe for this one particular application (as a precision NMR teslameter), the basic design appears to lend itself to other potential applications. We discuss several possibilities near the end of this article.

## 2  Discussion

### 2.1  Principles

Our probe is a saddle coil designed such that the coil itself constitutes a transmission line. A saddle coil is similar to a Helmholtz coil, but with the conductors constrained to lie on the surface of a cylinder [1]. Unlike conventionally tuned probes, which rely on matching networks to impedance match the probe to the spectrometer, a properly terminated transmission line probe presents a constant impedance to the spectrometer, independent of RF frequency. Furthermore, as it contains no resonant circuit, a transmission line probe is inherently equally sensitive to all NMR signals within its bandwidth, and requires no tuning to move between frequencies of interest. These two characteristics make transmission line probes fundamentally different from other types of NMR probes which employ transmission lines. In solid state NMR for example, there are many designs for multiply tuned cross-polarization magic angle spinning probes which use transmission lines as tuning and matching elements [2]. In our design, however, the probe itself is a transmission line: no tuning or matching is required. Note that this is also distinct from the situation in which a transmission line with a carefully chosen electrical length is attached to a probe in order to move the tuning and matching elements farther from the coil.

The basics of transmission lines can be found in standard RF engineering texts, e.g. Pozar [3]. A key feature to recall in the context of this work is that the characteristic impedance of a transmission line is determined purely by its geometry and by the materials involved in its construction. When terminated with a load which matches its characteristic impedance, no reflections will appear on the line. In a loss-free transmission line with uniformly distributed inductance and capacitance, the impedance is independent of frequency and is given by:

$$Z_0 = \sqrt{\frac{L/l}{C/l}}, \qquad (1)$$

where $L/l$ is the inductance per unit length and $C/l$ is the capacitance per unit length. The basic premise behind a transmission line NMR probe then is to match the distributed inductance, which appears due to the probe coils, with an appropriate distributed capacitance, in order to give the desired characteristic impedance. The line can then be terminated with a resistor of the same value. The desired characteristic impedance is determined by the spectrometer, and in most cases is 50Ω.

## 2.2 Previous Work

The use of transmission line-like probes in NMR was originally discussed by Lowe, Engelsberg, and Whitson [4,5] in the mid 70's, then recently revisited by Kubo and Ichikawa [6]. A significant difference between these authors' work and our own is that their probes are modeled as delay lines consisting of a finite number of cascaded LC circuit sections. This is natural for their designs, because their probes are constructed with a small number of discrete capacitors and wire inductors arranged in segments, forming a classic delay line. These probes are thus described by a characteristic impedance

$$Z = Z_0 \sqrt{1 - \frac{\omega^2}{\omega_c^2}}, \qquad (2)$$

where $\omega_c$ is a cutoff frequency above which no signal will propagate. The cutoff frequency is determined by $\omega_c = \sqrt{\dfrac{1}{L_u C_u}}$, where $L_u$ and $C_u$ are the inductance and capacitance per circuit section.

In a transmission line, the inductance and capacitance are distributed continuously along the length of the line rather than appearing in discrete cascaded sections. There are then essentially an infinite number of these sections, each of infinitesimal length, with the important consequence that there is (in principle) no cutoff frequency. Since our probes have a continuously distributed capacitance and inductance (with no discrete elements used in the construction), they are best treated as transmission lines rather than delay lines. It is interesting to note that Kubo and Ichikawa [6] found that after including the mutual inductances between adjacent circuit segments, the impedance of discrete delay line probes could also be well described by the transmission line formula.

As mentioned by Kubo and Ichikawa [6], very few transmission line probes have been put into practical use, most likely due to the difficulty encountered until now in their construction. We have only found two examples of transmission line coils: that of Stokes [7], who used a solenoidal coil to measure the T1 of $^{19}$F in difluorotetrachloroethane over a range of 18-80MHz, and a low field ESR gaussmeter developed by Gebhardt and Dormann [8]. Both were solenoidal in design and required somewhat involved construction and calibration techniques. Our design constitutes a straightforward method of constructing transmission line probes. An important benefit of our method is that we can produce large quantities of practically identical probes with ease. Furthermore, our design is the first which uses a saddle coil, thus making it particularly appropriate for superconducting solenoidal magnets.

## 3   Design

### 3.1   Method

The basic idea behind our probe is to make a saddle coil pattern out of traces on a flexible circuit board. We model the probe inductance, which is dominated by the saddle coil loops, as being uniformly distributed along the length of the coil trace. Then, for a given coil size and number of loops, we control the capacitance per unit length by adjusting the trace width. The construction method is similar to that of planar surface microcoils [9], but because our circuit board is flexible, we can wrap it around a sample tube to create a saddle coil.

Calculating the characteristic impedance of our transmission line probe consists of two parts, the capacitance and the inductance. Because the size of our coils is small compared to the wavelength of interest, we calculate the electrical characteristics of our probe for DC fields. To design probes for use at higher frequencies, of course, these assumptions will break down and more sophisticated modeling will need to be employed. We chose trace widths large compared to the thickness of the flex substrate, so that the capacitance could be modeled as that of a simple parallel plate capacitor, with

$$C = \frac{\varepsilon_e A}{d} \qquad (3)$$

Here $A$ is the surface area of the entire probe trace, $d$ is the thickness of the dielectric, and $\varepsilon_e$ is its effective dielectric constant. The effective dielectric constant [10] combines the effects of the dielectrics above and below the transmission line. It is dependent on the trace width and thickness. A typical ratio of the effective dielectric to the material dielectric for our various probe geometries is 0.88 to 0.89 . To calculate the inductance of our coils, we initially made a simple estimate using the Biot-Savart law. We then used the results of Mohan et al. [11] for determining the inductance of surface-patterned spiral inductors. For various probe geometries the Mohan result was greater than the Biot-Savart approximation by around 18-20%.

To verify our calculations, we directly measure the total capacitance and inductance of our probes. To measure $C$, we remove the terminating resistor from the probe (so that the probe looks like an open circuit at DC) and read off the capacitance at the lowest frequency available (1 MHz) from a network analyzer. Using this method, our capacitance predictions match our measurements within 10-20% (see Table 1). To measure the inductance, we replace the resistor with a short circuit. These measurements, also made at 1MHz, match the predicted values within 20-40% (see Table 1).

## 3.2  Potential Pitfalls

We have considered two possible types of deviations from ideal behavior in our probes: impedance mismatches and non-uniform distribution of inductance and capacitance. Impedance mismatches can occur at the "launch lines" which connect the coaxial connector to the coil, as well as at the interface between the coils and the stripline which connects one coil to the other. Reflections at the coax to coil connection reduce the total power delivered to the coil, as well as signal received from it. Reflections at the coil to stripline interface can unbalance the currents in the two sides of the saddle coil, and hence introduce inhomogeneity in the RF field produced and detected by the probe [6]. Furthermore, reflections can also cause the current amplitude to vary along the length of the probe, which can introduce additional spatial variation to the field.

The second form of deviation from the ideal that our probes may experience is the non-uniform distribution of capacitance and inductance. In this case the theoretically infinite segmentation of the transmission line becomes effectively finite, and our probe will behave in a manner similar to the delay line probe discussed above, albeit with a potentially high number of circuit sections. This has two effects. The first is the appearance of a cutoff frequency, as described in Eq. [2]. The second is the appearance of a decay frequency, above which a signal will still propagate but will not effectively couple to the spins. This is due to phase delays between adjacent circuit segments [6]. The magnetic fields produced by the phase-delayed currents can interfere destructively, thus reducing the total RF field. The decay frequency is probe volume dependent; for cylindrical coils it scales as one over the radius cubed [6]. Since our probes have relatively large radii, if their inductance and capacitance are not uniformly distributed it is possible that this could be a limiting factor for higher frequencies.

## 4  Construction

The flex material we have used is the Pyralux LF brand by DuPont, which consists of 1 oz. copper (thickness = 34.2 μm) deposited on both sides of a Kapton polyimide dielectric substrate. For this copper thickness the dielectric is available in thicknesses ranging from 76 to 127 μm. The thinner varieties are susceptible to permanent damage by creasing (similar to tin foil), and care must be taken when handling them. A dielectric coverlay can be deposited over the final design in order to lend mechanical rigidity and electrical insulation.

We have constructed three different types of probe. The first (probe #1) was machined on a CNC mill, making use of a vacuum chuck to restrain the flexible circuit material during cuts. This probe has two loops, and is rolled to form a saddle coil one inch in diameter. It is fabricated on Pyralux LF 9151R substrate (dielectric thickness of 0.007", ε = 3.7). Difficulties in machining make this an unreliable method for producing multiple identical probes. However, this prototype probe showed excellent impedance characteristics and gave us confidence in our basic modeling.

A more convenient and reliable way of producing these probes is to have them fabricated by commercial flex circuit manufacturing houses. We have used two different manufacturers to produce three different types of probe. The first (probe #2) is a one inch diameter, three loop probe on Pyralux LF 7010R, with trace widths of 0.042". It has a 0.002" thick dielectric (Kapton + adhesive) with a dielectric constant $\varepsilon = 3.7$. The second (probe #3) is a one inch diameter two loop probe on a non-specified 1oz. copper laminate, with trace widths of 0.074" and a dielectric with a thickness of 0.007" and $\varepsilon = 3.5$. Our most recent probe design (probe #4) is a 15mm diameter two loop probe printed on Pyralux LF 9151R, with trace widths of 0.062" and a dielectric thickness of 0.007", $\varepsilon = 3.7$. The coverlay for probes 2-4 is LF 1510, with thickness 0.0015". Connections to and from the probe are made via a U.FL standard miniature (0.81mm diam.) coaxial cable produced by Hirose. The transmission line is terminated on the flex circuit, with a 1/4 Watt 50 ohm non-magnetic chip resistor from SRT Technologies (model CHR-1206). The straight lines of conductor between the coaxial launch and the coils are standard 50Ω microstrip transmission lines. A photograph and mechanical drawing of probe #4 can be found in Figs. 1 and 2. Once manufactured, the probe is secured via plastic cable ties around a tube appropriate for the sample of choice.

Because of the limited thicknesses of dielectric available, the capacitance per unit length of the microstrip can be designed only over a fairly restricted range. This limitation can cause difficulties when attempting to specify the value of $C/l$ needed to match the large $L/l$ of the coil loops as necessary to make a $Z_0 = 50\Omega$ line. We found that trying to meet this capacitance requirement effectively limited us to no more than three loops.

## 5  Results

A natural measure for an initial evaluation of the electrical properties of our probes is the reflection coefficient $\Gamma$, defined as the ratio of the reflected voltage to the incident voltage:

$$\Gamma = \frac{V_{ref}}{V_{inc}}. \tag{3}$$

This quantity is easily measured with modern network analyzers and is simple to interpret. In particular, this coefficient determines the fraction of power delivered to the load:

$$P = P_0(1 - |\Gamma|^2). \tag{4}$$

Here $P_0$ is the incident power and $P$ is the power delivered to the load. In our situation the load is either the probe or the receiver amplifier, depending on whether the spectrometer is transmitting or receiving.

The transmitted power as a function of frequency for probe #4 can be seen in Fig. 3. The measurements were taken with a Hewlett-Packard 8712C Network Analyzer. As seen in Fig. 3, the probe is extremely well matched to a 50$\Omega$ load over the frequency range of 5 to 26 MHz (0.1 to 0.6 T) required for our experiment; essentially all of the applied power is transmitted to the load. At higher frequencies, however, the probe begins to reflect larger fractions of the power. At 150MHz nearly 90% of the applied power is transmitted, but at 300MHz almost half of it is reflected.

Having a large transmission coefficient does not ensure that power is delivered in the form of an RF magnetic field, however.  Field strength can be affected either by the decay frequency effects discussed above, or by current being shunted through the capacitance of the probe instead of flowing through the coils as desired.  Hence, a more stringent test of our probes requires measuring the RF field produced by the probe as a function of both frequency and position.  To do this, we made a small pickup coil of one ~1.2mm radius loop terminated with a 50$\Omega$ resistor.  In separate measurements, we verified that this pickup coil has a useful bandwidth of DC-300MHz.  To measure the RF field at various positions in the probe, we mounted the pickup coil on a translation stage near the center of a probe.  The probe was driven by a function generator, and the voltage induced in the pickup coil was amplified and monitored with an oscilloscope.  The RF field from the probe was investigated at frequencies up to 300 MHz using this setup.   At low frequencies, the behavior of the probes agreed well with our expectations based on modeling with DC currents.  At higher frequencies, substantial changes appeared both in the spatial distribution of the RF field, and in its magnitude.  Figs. 4 and 5 show this behavior.

Fig. 4 shows the measured RF field strength, as a function of position along the axis which connects the saddle coil centers, for various frequencies.  At low frequencies, we expect a field minimum halfway between the coils, as observed.  As the frequency is increased, the field minimum shifts away from the center, and towards the second coil in the probe (the coil furthest from the launch connector, and closest to the terminating resistor). At 300MHz, the field minimum shifted more than a millimeter from the center. Similar data showed that, in contrast, the RF field distribution along the central axis of the saddle coil cylinder (perpendicular to the line between coil centers) was relatively insensitive to the frequency. This indicates that the reason for the change in transverse field distribution is simply that the field produced by the first coil is stronger than that of the second. We suspect that this may be due to reflections at the transitions from the coils themselves to the thin microstrip trace connecting the coils.

Results for the absolute RF field strength as a function of frequency appear in Fig. 5. To account for the varying spatial distribution of the field, this measurement was performed while tracking the field minimum along the axis connecting the coils. The measured values take into account the measured amplifier gain vs. frequency, and include corrections for the measured pickup coil behavior as a function of frequency. We compare the measured values with predictions based on the calculated low-frequency behavior. The predictions include corrections for the measured reflection coefficient of the probe and the measured transmission properties of the U.FL cable. The measured and predicted values agree to within 25% for all frequencies up to 150MHz, but begin to diverge substantially at higher frequencies. Considering that the probe was designed using electro- and magnetostatic formulae, the frequency range of good performance is surprisingly broad. Several possible reasons for deviation from the predicted behavior were mentioned earlier; however, determination of the source of the degradation in performance at high frequencies will require further investigation.

In order to further characterize the RF field homogeneity of our probe, we used the 4T magnet at the Yale Magnetic Resonance Research Center to make a gradient echo image of probe #4's RF field at 170 MHz. As seen in Fig. 6, the probe field is unbalanced, with greater sensitivity near one coil and less near the other. The overall behavior is consistent with our pickup-coil measurements. Both techniques indicate that at frequencies above about 100 MHz, the expected region of RF field homogeneity near the center of the saddle coils shifts to one side and becomes narrower. We note in passing that using the probe with a commercial Bruker console required no extra consideration beyond a cable adaptor.

We have tested the NMR performance of our probes in an Oxford 85/310HR horizontal superconducting magnet run at a proton frequency near 22MHz. The field was shimmed to approximately 0.1ppm inside a 5cm long, 15mm diameter cylinder along the bore. The spectrometer is home built and will be described elsewhere. The important characteristic in terms of its RF performance is that we have replaced the traditional $\lambda/4$ line and crossed-diode based protection circuit with a simple electromechanical reed relay (Fig. 7). At the center of the magnet and using a long, cylindrical water sample doped with 68.5mM $CuSO_4$, we can measure the proton frequency with a typical standard deviation in the center frequency of around 0.4Hz for data from a single free induction decay. Under these conditions the linewidths for probe #4 are around 12.5Hz, while the linewidths for probe #3 are around 25Hz. After switching to pure water we see linewidths of around 6Hz for probe #4. We believe the residual line widths are a result of the inhomogeneous B0 field; this is consistent with the fact that the measured linewidth for probe #4 is smaller than that of probe #3, due to the smaller volume sampled by probe #4.

In order to verify the broadband performance of our probe, we obtained spectra from [1]H (in $H_2O$ and in $H_2PO_4$), [19]F ( in perfluorotributylamine [Fluorinert, FC-43]) and [31]P (in $H_2PO_4$). Examples can be seen in Fig. 8. Switching between the various nuclei required changing only the samples and the applied RF frequency.

# 6   Possible Improvements

We anticipate that this probe might be useful in a wider range of circumstances than in our experiment, so here we discuss some possible improvements in the design. Clearly, operation up to higher frequencies would be desirable for many applications. A first step in this direction would be simply to make the probes smaller. Making the characteristic length of the probe smaller may reduce the effects of reflections. Evidence for this can be seen in Fig. 9: The useful frequency range (defined loosely as frequencies below the first peak in the voltage reflection coefficient) for the 15mm diameter probe #4 is significantly broader than that of the 25.4mm diameter probe #3. Furthermore, if the capacitance and inductance are unevenly distributed along the probe, thus causing it to behave more like a delay line rather than a transmission line, then one would expect the phase delays between sections to decrease thus raising the decay frequency. Finally, additional improvements in coil design may be possible by more accurately predicting the probe's behavior at higher frequencies (e.g. by using commercial RF electromagnetic simulation packages).

# 7  Potential Applications

The utility of a transmission line probe goes well beyond that of our particular application (a high-precision, broadband NMR teslameter for a physics experiment). Other potential uses include:

- The ability to simultaneously acquire the spectra of several nuclei without the need for multiply tuned probes.

- Use in solid-state NMR, where the spectral frequency ranges are often too large for conventional tuned probes [6].

- NMR of nuclei with large quadrupolar couplings, where frequency sweeps are typically employed [6].

- Study of samples whose magnetic properties are dependent on the external field strength, since with a non-resonant probe one can change the strength of the static field continuously [6].

- Use in cryogenic studies to avoid tuning difficulties due to resonant circuit components at different temperatures [12].

- Use in short time scale experiments, since non-resonant probes should have better transient recovery times [4].

- Reduction in Johnson noise by locating the terminating resistor in a low temperature bath [6].

- Use in nuclear quadrupole resonance detection schemes, e.g. for explosives and narcotics, where the signals intrinsically cover a broad range of frequencies [13] and fast transient recovery time can increase the signal to noise for fast decaying-signals [3, 13].

## 8   Conclusion

We have designed and constructed an easy to manufacture, tuning free, broadband NMR probe.  The probe coil itself constitutes a transmission line, which ensures a good impedance match to a standard spectrometer across over 200 MHz of bandwidth. Construction of our probe is easily performed by printing the coil design onto a flexible circuit board, and design of new probe geometries should be straightforward.

# 9 Acknowledgements

We thank R. Schoelkopf, R. Vijayaraghavan, E. K. Paulson and B. Turek for many useful discussions. We also thank K. M. Koch and R. A. De Graaf for making the RF field map of probe #4. Additionally we thank E. K. Paulson for providing us with the Fluorinert sample.

This material is based upon work supported by the National Science Foundation under Grant No. 0457039.

**Table 1:** Electrical and mechanical parameters for three probe designs. Experimental values were measured at 1MHz, predicted values are at DC. The diameter refers to the total probe diameter, or the distance between the centers of the coils when the probe is wrapped around a sample tube.

| | Diam. | # loops | $C_{meas}$ | $L_{meas}$ | $\dfrac{C_{meas}}{C_{pred}}$ | $\dfrac{L_{meas}}{L_{pred}}$ | $Z_0 = \sqrt{\dfrac{L_{meas}}{C_{meas}}}$ |
|---|---|---|---|---|---|---|---|
| Probe #2 | 0.5" | 3 | 375pF | 769nH | 0.86 | 0.76 | 45.3 |
| Probe #3 | 0.5" | 2 | 138pF | 343nH | 1.02 | 0.95 | 49.8 |
| Probe #4 | 15mm | 2 | 69pF | 230nH | 1.10 | 1.43 | 57.7 |

**Figure 1:**     Photograph of probe #4.  The U.FL RF connector is located outside the picture, at the end of the microstrip "launch line" extending down past the bottom of the picture.

**Figure 2:**     Mechanical drawing of front side of probe #4.  The circuit is wrapped around a sample tube to form a volume coil that is 15mm in diameter and 15mm long.  Trace widths are 0.062".  A ground conductor is patterned on the back underneath the traces.  Connections between the front and back are made via 0.007" diameter circular plated through holes.  The T shaped lines are where the terminating resistor sits.  The coaxial connector is on a tail several inches away, outside the drawing.  The long "tail" was incorporated into the design when we found that the U.FL connectors contain a nickel flash of sufficient volume to affect our NMR measurements when located too near the probe.

**Figure 3:**     Measurement of the unreflected fraction of power, $(1 - |\Gamma|^2)$, vs. frequency, for probe #4.  $\Gamma \equiv V_{reflected} / V_{incident}$.

**Figure 4:**     RF field strength versus position for probe #4 along the axis connecting the centers of the coils.  Measured results are shown at several different frequencies, as well as the calculated field profile at DC.  The zero of the x-axis is defined as the fitted position of the field minimum at low frequency.  To highlight the changes in the spatial distribution of the field, results from all frequencies are normalized to the same value at x=0.  A shift in the position of the field minimum, as well an increase in the spatial curvature of the field, is evident at high frequencies.

**Figure 5:**     Measured and predicted RF field strength vs. frequency for probe #4, at an input power of +16dBm (40mW).  As discussed in the text, the field strength for each frequency is measured at the position corresponding to the local minimum near the center of the saddle coils. The absolute accuracy of these measurements is at the 10-20% level.  The predicted field strength is based on magnetostatic calculations, with corrections for the measured voltage reflection from the probe.

**Figure 6:**     Gradient echo image of the RF field produced by probe #4 at 170MHz.  The coils are near the upper left and lower right corners.  This is a nutation angle map made from a 3.0mm axial slice taken near the longitudinal center of the probe.  TR = 3s; TE = 3ms; 2 Averages; Spectral Width = 50 kHz; FOV = 1.6 x 1.6 cm, 50 x 50 pixels.

**Figure 7:**     Broadband duplexer.  Relay is Coto 2341.  The rated switching time is 0.5 ms and the carry current rating is 1.5 A.  The frequency range of this relay is unspecified, but it performs well below 22MHz.   Relays with similar specifications, specifically designed to act as 50 $\Omega$ transmission lines for signals with RF frequencies up to the GHz range, are available if needed.

**Figure 8:**     Sequentially obtained spectra from various nuclei.  The (a) $^1$H  and  (c) $^{31}$P spectra come from $H_2P0_4$ (85% by mass); the (b) $^{19}$F spectrum comes from perfluorotributylamine (Fluorinert, FC-43).  Samples were held in a 17x120 mm standard 15mL Falcon plastic tube.  Both  the $^1$H and $^{31}$P spectra are the average of 100 FID's, each of 150,000 samples with a 2s delay between each FID acquisition.  The $^{19}$F spectrum is the average of 100 FID's, each of 45,000 samples, with a 2s delay between each FID acquisition.  In all cases the sampling rate was 1MHz.  The only adjustment made to the spectrometer between acquiring these signals was a change in the drive frequency and the tipping pulse time.  The tipping pulse was 80μs for $^1$H and $^{19}$F, 200μs for $^{31}$P.

**Figure 9:**      Voltage reflection coefficient vs. frequency for probes #3 and #4.  The smaller size of probe #4 pushes the peak in reflected power to a higher frequency.

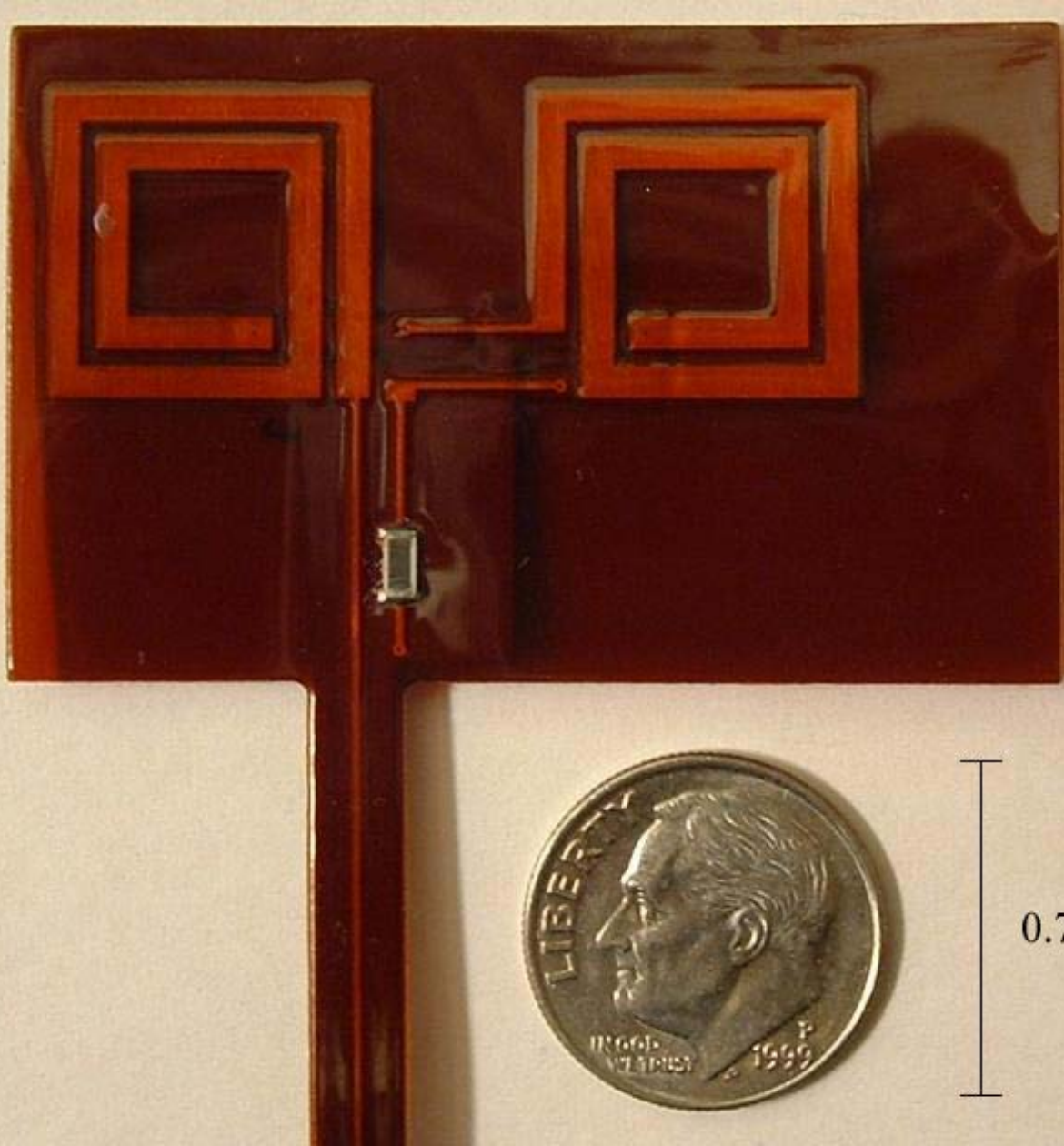

0.7"

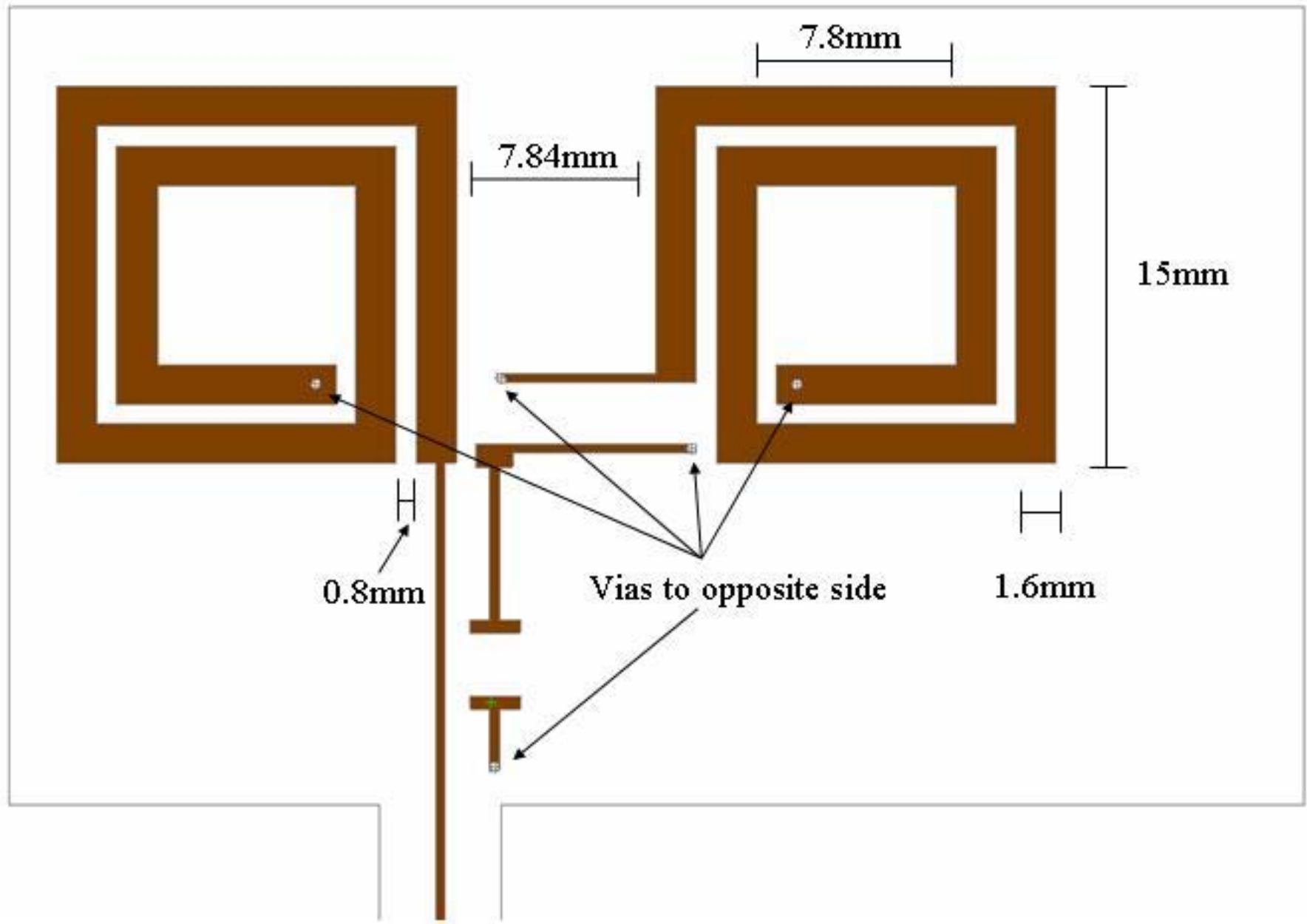

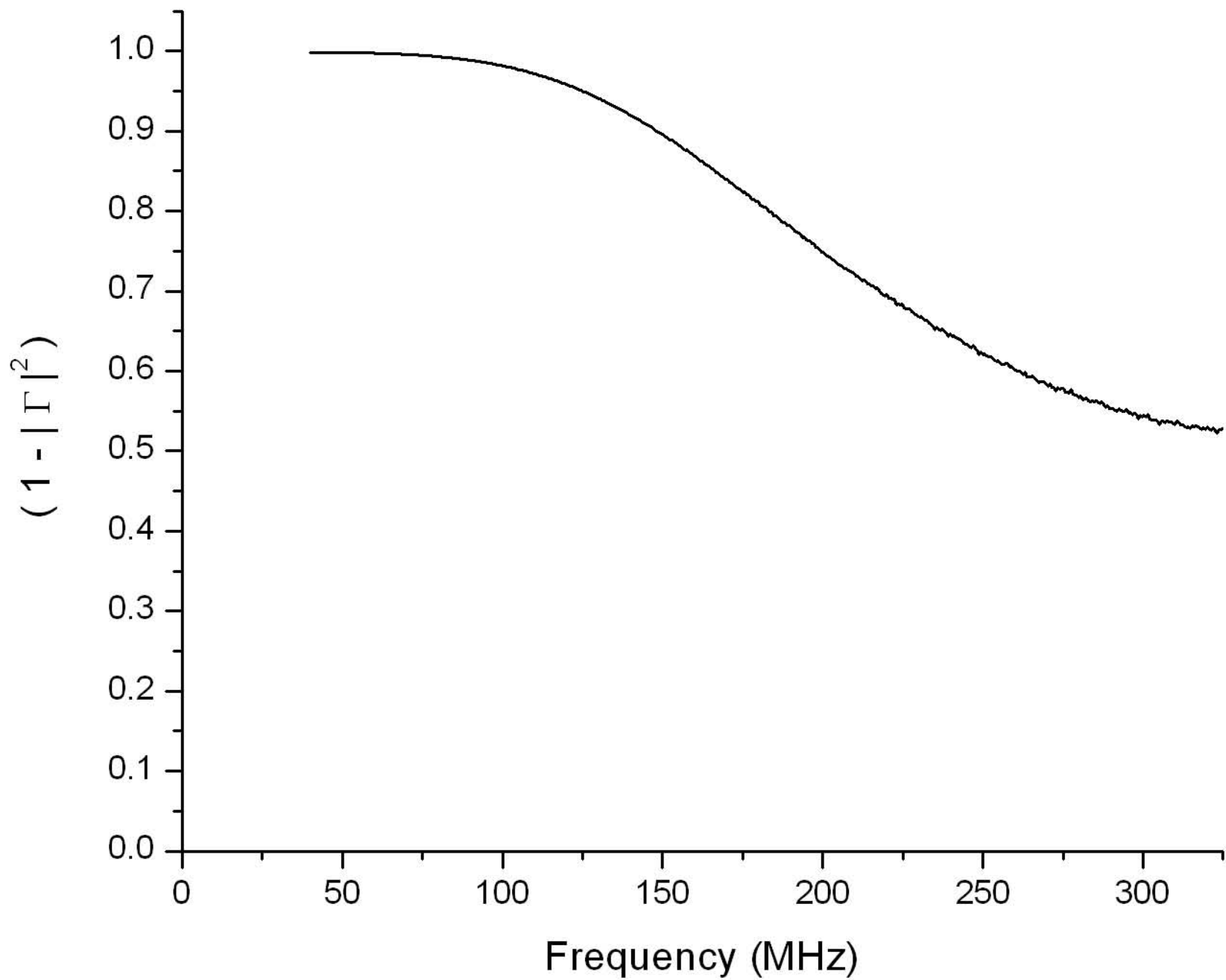

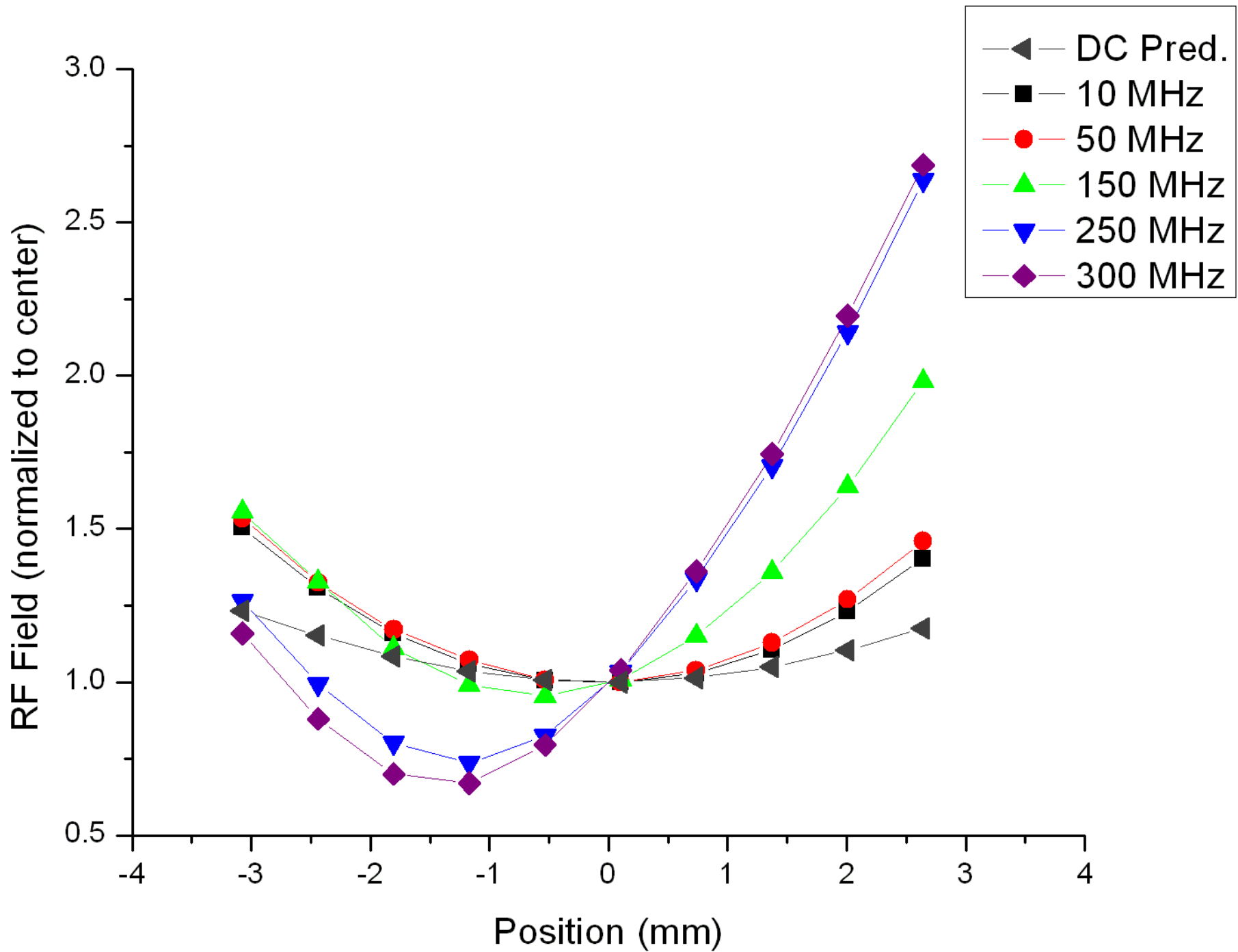

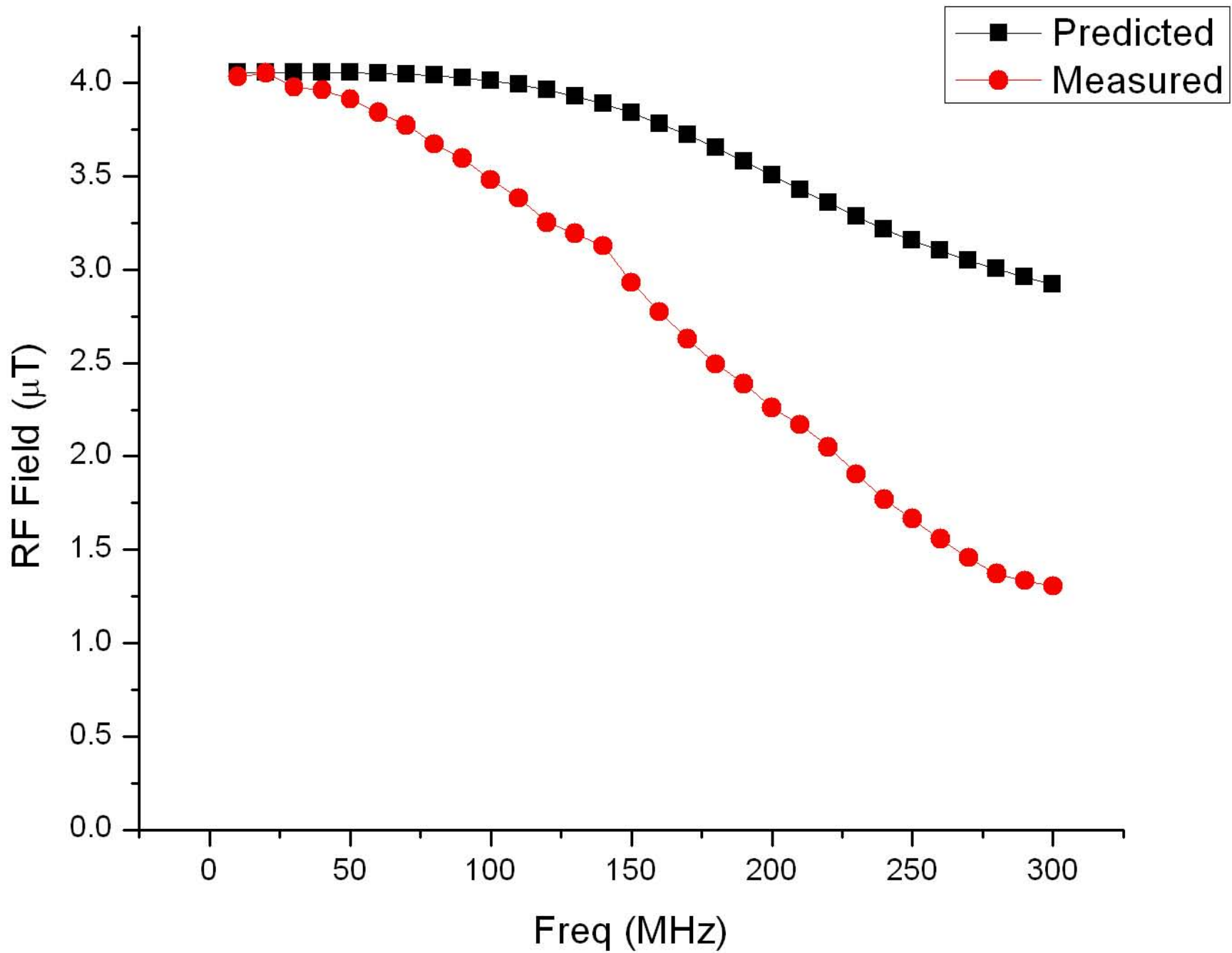

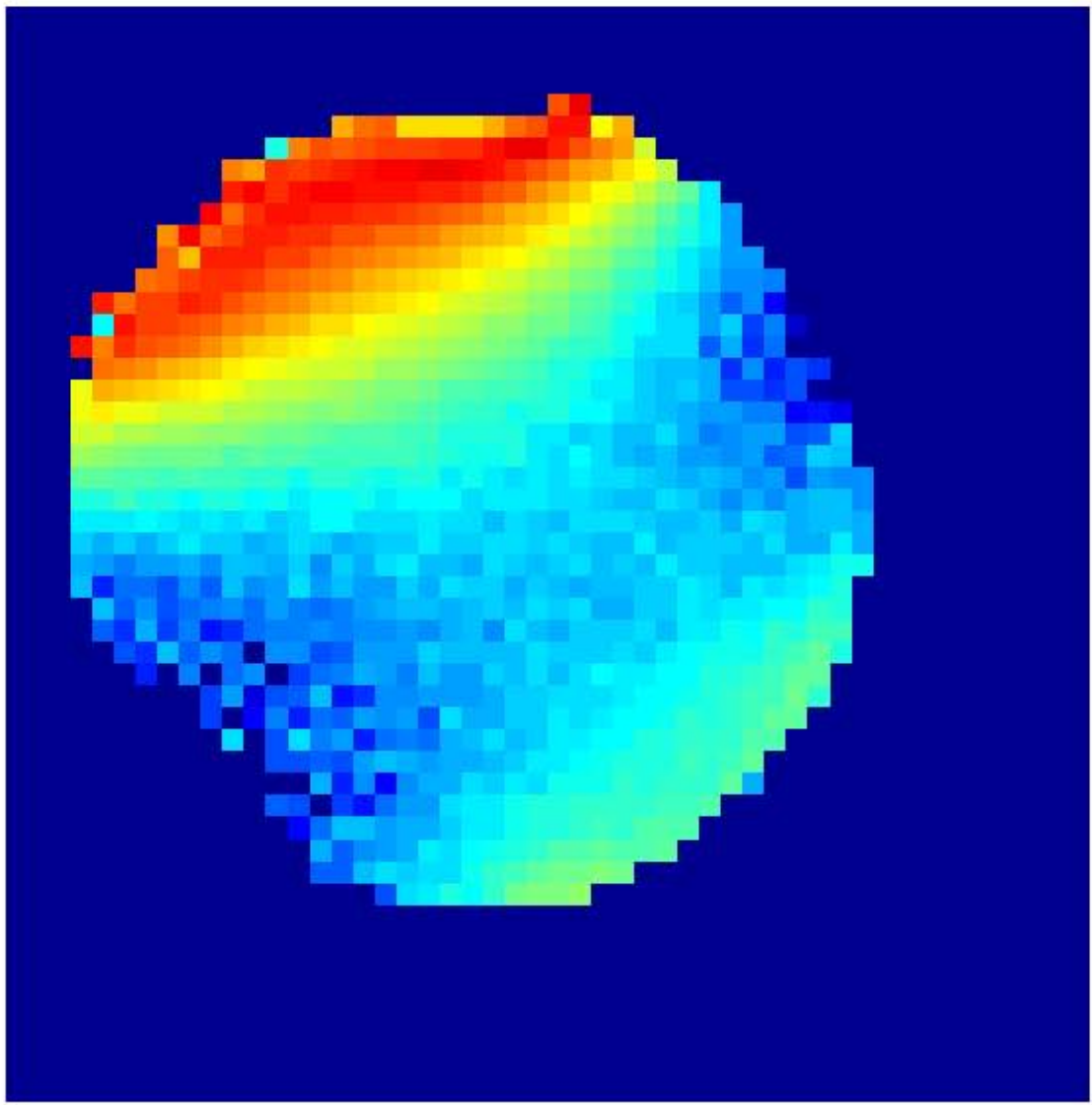
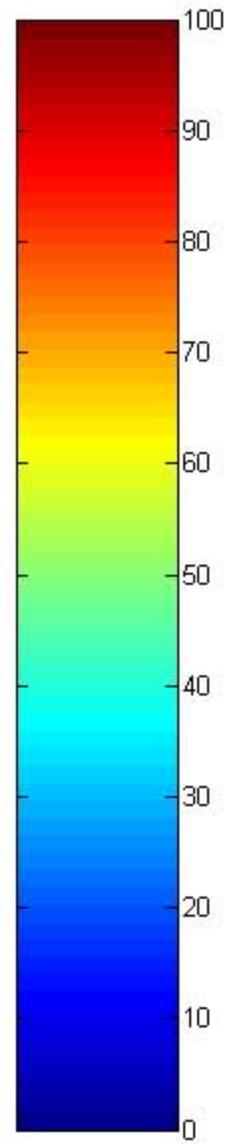

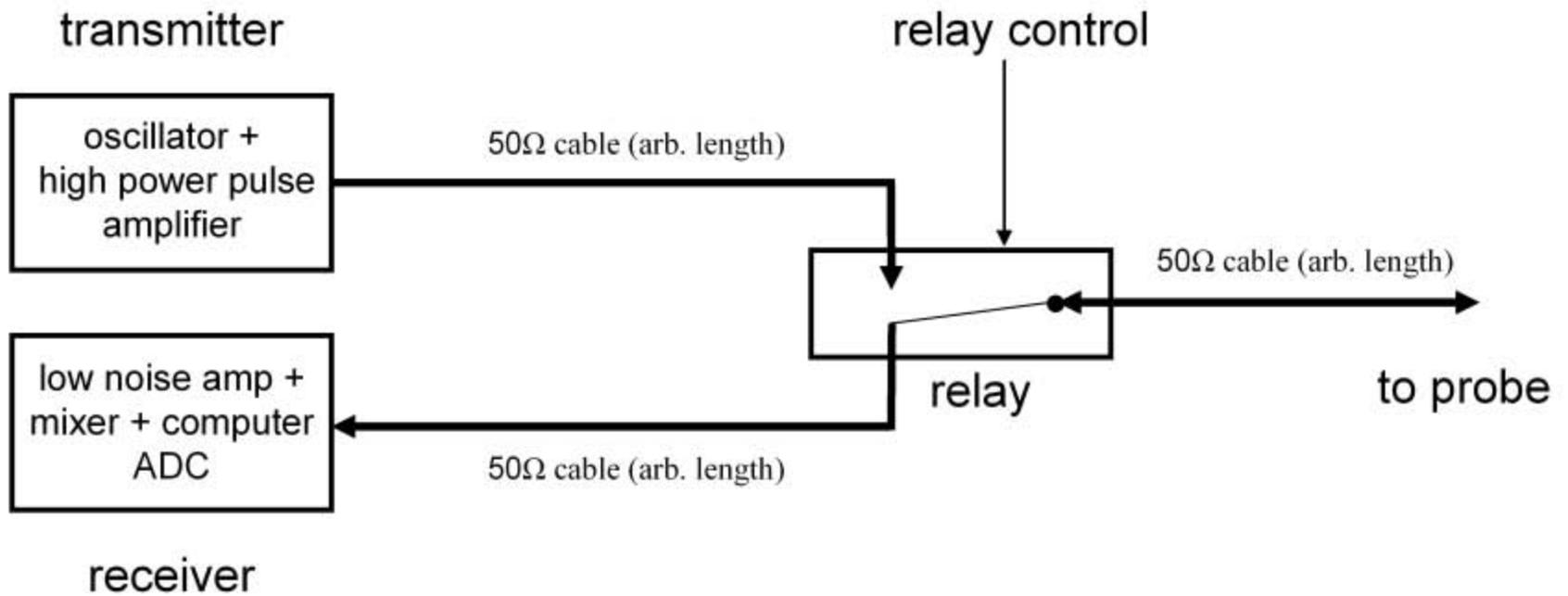

**transmitter**

oscillator +
high power pulse
amplifier

50Ω cable (arb. length)

**relay control**

50Ω cable (arb. length)

**relay**

**to probe**

low noise amp +
mixer + computer
ADC

50Ω cable (arb. length)

**receiver**

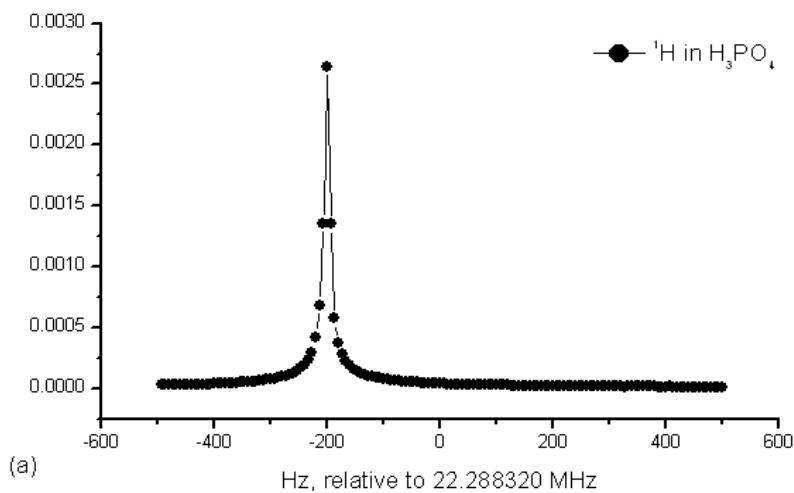

(a) Hz, relative to 22.288320 MHz

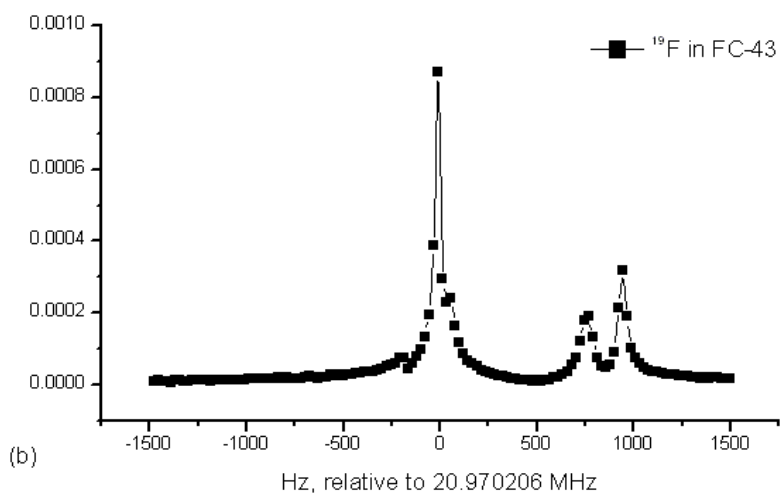

(b) Hz, relative to 20.970206 MHz

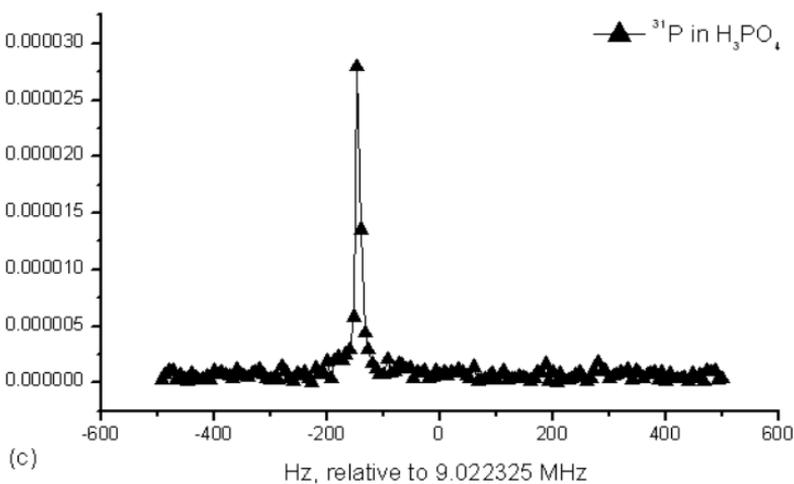

(c) Hz, relative to 9.022325 MHz

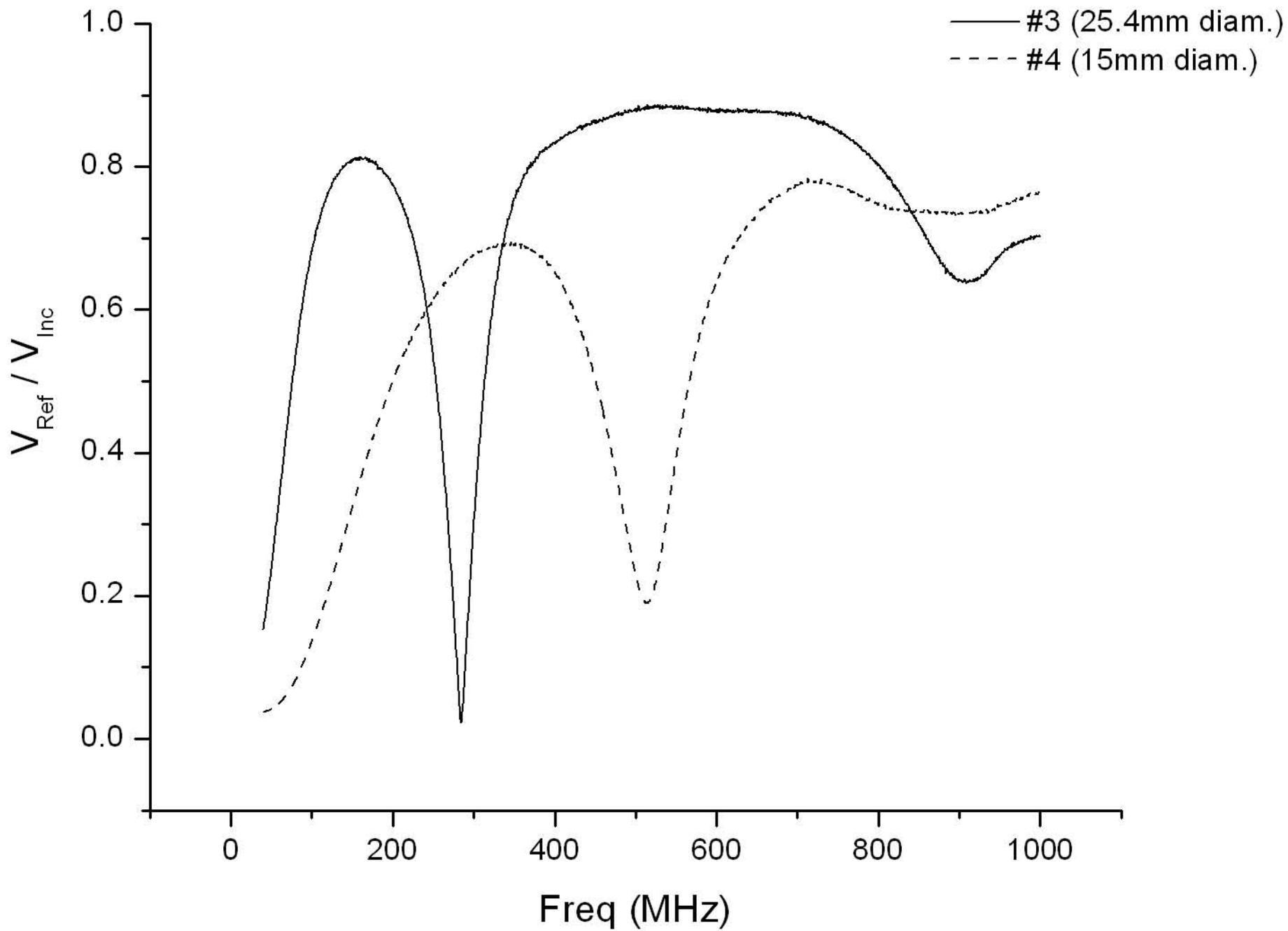